# Mapping the nanomechanical properties of graphene suspended on silica nanoparticles

Z. Osváth[a,c,*], E. Gergely-Fülöp[a], A. Deák[a], C. Hwang[b,c], and L. P. Biró[a,c]

[a]*Institute of Technical Physics and Materials Science (MFA), Centre for Energy Research, HAS, 1525 Budapest, P.O. Box 49, Hungary*

[b]*Center for Nano-metrology, Korea Research Institute of Standards and Science, Yuseong, Daejeon 305-340, South Korea*

[c]*Korea-Hungary Joint Laboratory for Nanosciences (KHJLN), 1525 Budapest, P.O. Box 49, Hungary*

**Abstract:** Using nanoparticles to impart extrinsic rippling in graphene is a relatively new method to induce strain and to tailor the properties of graphene. Here we study the structure and elastic properties of graphene grown by chemical vapour deposition and transferred onto a continuous layer of $SiO_2$ nanoparticles with diameters of around 25 nm, prepared by Langmuir-Blodgett technique on Si substrate. We show that the transferred graphene follows only roughly the morphology induced by nanoparticles. The graphene membrane parts bridging the nanoparticles are suspended and their adhesion to the AFM tip is larger compared to that of supported graphene parts. These suspended graphene regions can be deformed with forces of the order of 10 nN. The elastic modulus of graphene was determined from indentation measurements performed on suspended membrane regions with diameters in the 100 nm range.

[*]Corresponding author: Zoltán Osváth, e-mail: zoltan.osvath@energia.mta.hu.

**Introduction**

Graphene has exceptional mechanical properties like the extremely high in-plane stiffness and superior strength [1, 2]. These properties are very important for applications in nanoelectromechanical systems, or to fabricate nanocomposites with graphene inclusions as structural material. The elastic constants of graphene obtained by various calculation methods are summarized in a recent review [3]. The Young's modulus of monolayer [4] and few-layer graphene [5] was measured by AFM indentation on micrometre-scale suspended graphene, bridging trenches or wells etched in silicon dioxide. It was experimentally found that the graphene shows both non-linear elastic behaviour and brittle fracture [4]. Here, we were able to study the elastic properties of graphene suspended between $SiO_2$ nanoparticles (NPs), by deforming nanomembranes with diameters in the range of 100 nm, i.e. one order of magnitude less than in previous experiments [4, 6]. We mapped by PeakForce® atomic force microscopy (AFM) the nanoscale mechanical properties of graphene transferred onto a continuous layer of $SiO_2$ NPs. We showed the differences in adhesion and deformation characteristics between supported and suspended graphene regions by comparing the nanomechanical data with topography. Based on the acquired adhesion maps we identify extended graphene regions suspended between NPs and show that these suspended regions can be deformed by forces of the order of 10 nN. Additionally, by transferring graphene onto $SiO_2$ NPs we introduced significant extrinsic rippling in graphene [7, 8]. Such rippling can affect the electronic properties of graphene by contributing to the scattering of charge carriers [9, 10].

Nevertheless, rippled graphene can be good candidate for sensing, as simulations [11, 12] predict enhanced chemical activity in corrugated graphene.

**Materials and Methods**

Amorphous silica NPs with diameter of ~25 nm were synthesized according to the Stöber-method [13, 14]. First, a solution containing 50 ml ethanol, 1.594 ml of $NH_3$ (32 %) and 0.44625 ml of ultrapure $H_2O$ was stirred for 30 minutes. Then, 2 ml tetraethyl orthosilicate (98 %) was added to this solution and stirred overnight. Finally the ammonia was removed by distillation at 60 °C. Langmuir-Blodgett (LB) films of the nanoparticles were prepared in a KSV 2000 film balance. The ethanolic solution of NPs was sonicated for 5 minutes, then mixed with chloroform (Scharlau, reagent grade, stabilized with ethanol) and spread at the air/water interface. The particles were compressed at a barrier speed of 0.4 $cm^2/s$. When the surface pressure reached ~1 mN/m, the speed was lowered to 0.2 $cm^2/s$. The LB films were prepared by vertical deposition (6 mm/min) at ca. 80 % of the collapse pressure, which was measured before. We used silicon wafers as substrates, cleaned with acetone, water, 2 % hydrofluoric acid solution, and finally rinsed in water.

Graphene was grown by chemical vapour deposition (CVD) on a 25 μm thick copper foil (99.8% purity). The CVD furnace was evacuated to ~$10^{-4}$ torr and the temperature was raised to 1010 °C with $H_2$ gas flow (~$10^{-2}$ torr). When the temperature became stable, both $CH_4$ (20 sccm) and $H_2$ (5 sccm) were injected into the furnace for 8 minutes to synthesize the graphene. After the growth, we cooled down the furnace with a cooling rate of 50 °C/min.

The transfer of graphene onto the $SiO_2$ NPs was done using thermal release tape (TRT, Graphene Supermarket) and copper etchant (20% of $CuCl_2$ mixed with 37% HCl in 4:1 volume ratio). After etching the copper, the TRT holding the graphene was rinsed in distilled water, dried and pressed onto the nanoparticle-decorated Si surface. The sample was placed on a hot plate heated above the release temperature of the tape (90 $^{o}$C). After 1 minute the TRT was removed, leaving behind the graphene on top of $SiO_2$ NPs. The sample was annealed at 400 $^{o}$C in $N_2$ atmosphere for 2 hours in order to improve the contact of graphene with the NPs.

The sample was investigated both before and after annealing by a MultiMode 8 AFM from Bruker operating under ambient conditions. PeakForce® tapping mode was used to perform scans at well-defined forces. Peak Force tapping is a relatively new scanning mode, where the probe and sample are intermittently brought together (similar to Tapping Mode) to contact the surface for a short period, which eliminates lateral forces. While in Tapping Mode the feedback loop keeps the cantilever vibration amplitude constant, Peak Force tapping controls the maximum force (Peak Force) on the tip [15]. A complete force-distance curve is performed in every measuring point, while the *z*-piezo data of the cantilever is recorded at the maximum force between the sample and the cantilever. Analysis of force curve data is done on the fly, providing maps of multiple mechanical properties (e.g. adhesion, deformation, etc.) that have the same resolution as the height image. The maximum force can be changed in order to acquire images at different sample-cantilever forces. For the indentation experiments on suspended graphene, we used an AFM cantilever with tip radius R $\simeq$ 8 nm and spring constant k = 10.4 N/m, as determined *in situ* by the thermal tune method [16].

**Results and Discussion**

During the dry transfer procedure using thermal release tape, the CVD graphene breaks into sheets with different sizes, typically of several micrometres. The AFM topographic image in Figure 1a shows the film of closely spaced $SiO_2$ NPs, and a graphene sheet on top of NPs.

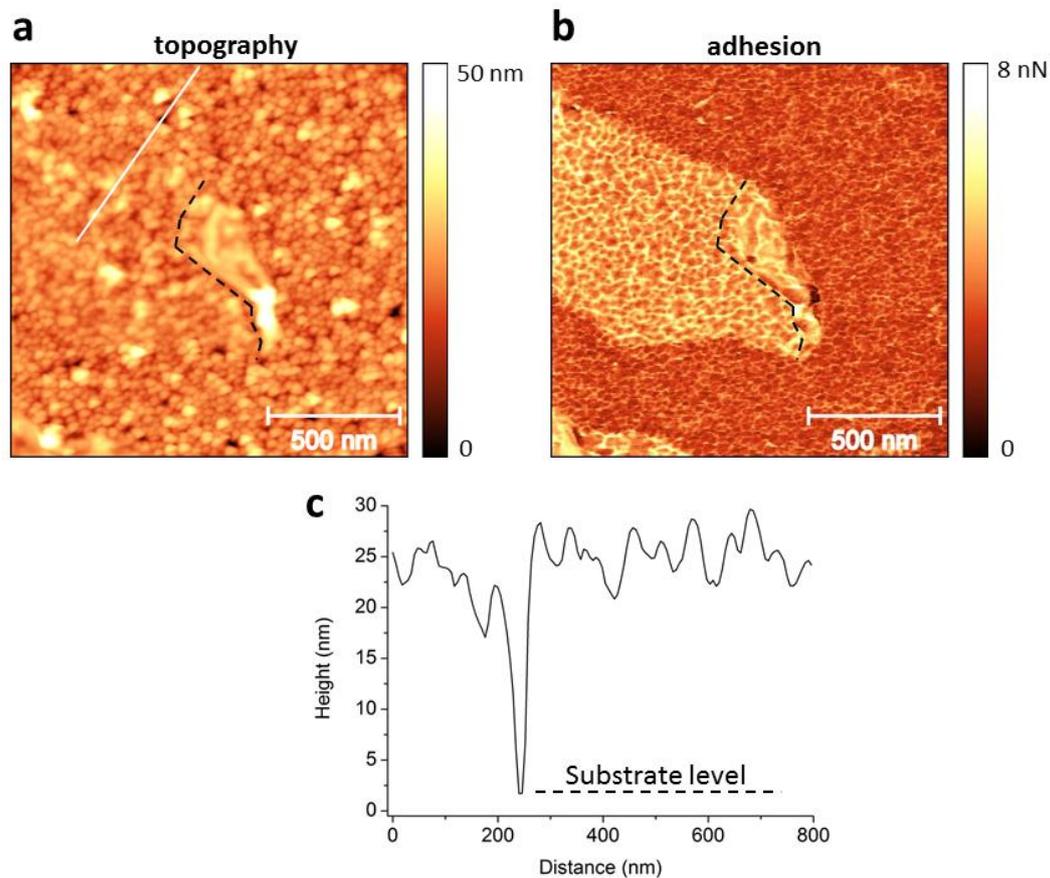

**Figure 1:** Peak Force AFM images of graphene-covered $SiO_2$ nanoparticle film. (a) AFM topographic data. (b) Adhesion map corresponding to the area in (a). (c) Line section along the white line in (a).

Since the graphene follows the corrugation induced by the NPs it is hardly observable in the topographic image. The measurements were performed in PeakForce® mode, where the adhesion properties of the surface can be recorded in parallel with topographic data. Figure 1b shows the adhesion map of the area presented in Figure 1a. The light-coloured regions show increased adhesion between the sample and the AFM tip and these regions correspond to graphene-covered areas, whereas dark-coloured regions correspond to bare $SiO_2$ NPs. A line section taken along the white line in Figure 1a is shown in Figure 1c, which shows the corrugated nature of graphene. This line section was drawn intentionally to include pit, i.e. a dark-coloured spot on the topographic image. This pit corresponds to a small area on the Si substrate where no NPs are present, which allows measuring the diameters of the NPs.

According to Figure 1c, the average height of the nanoparticles relative to the Si substrate is around 25 nm, which approximately corresponds to an average nanoparticle diameter. The line section also shows that graphene is completely separated from the Si substrate. It is either supported by NPs, either suspended between them, but it never reaches the substrate level. Note that the end part of the graphene sheet is back-folded, and this folded region is easier to observe in the topographic image (centre of Figure 1a). We marked with dashed line in Figure 1a-b the border of this back-folded graphene.

We acquired AFM images at different tip-sample interaction forces. For example, Figure 2a and 2b show the topographic images of the same graphene-covered area measured with tapping forces of 1 nN and 40 nN, respectively.

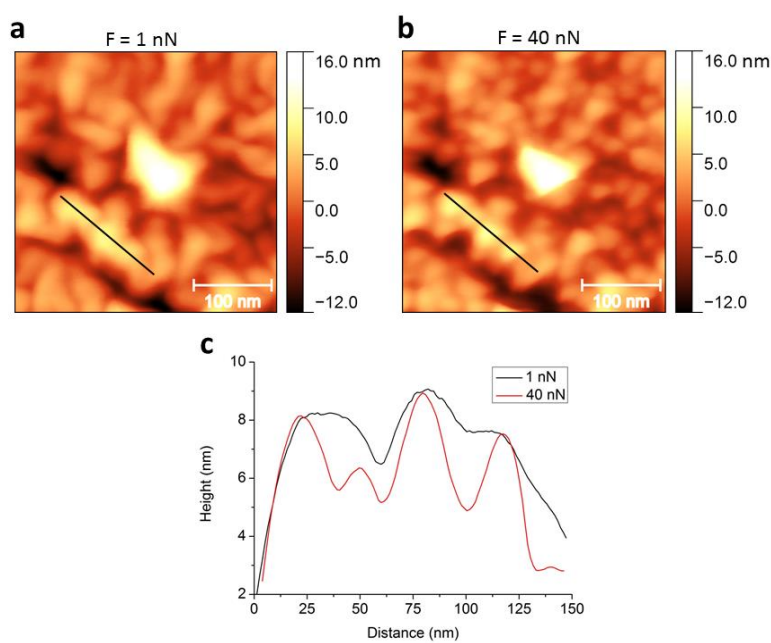

**Figure 2:** Peak Force AFM images of $SiO_2$ nanoparticle-supported graphene measured at a tip-sample force of (a) 1 nN and (b) 40 nN. (c) Height profiles along the line sections in (a)-(b) (black lines).

One can observe that the measured topography is slightly different in these two cases. To illustrate this, let us consider the two height profiles in Figure 2c, taken along the same line section (black line in Figure 2a-b), corresponding to the two different tapping forces.

The height profile measured at 40 nN displays four peaks, while the profile at 1 nN displays only three. Additionally, at 1 nN the height values are considerably larger between the peaks. This shows that graphene is actually loosely bound to the nanoparticles, not following exactly their morphology. There are significant suspended parts between the NPs, and the low force used for the measurement does not deform the structure. In contrast, when imaged at 40 nN, these suspended parts are deformed and pushed against the underlying NPs. Hence, a fourth (smaller) peak is evidenced in the height profile in Figure 2c, which corresponds to a nanoparticle with smaller diameter. Comparing Figure 2a and Figure 2b, we can assert that in general, the NPs under graphene are better resolved when imaging at higher tip-sample forces. The light-coloured area in the centre of Figure 2a-b is attributed to a small cluster of nanoparticles which protrudes from the otherwise monolayer NP film.

Next, we annealed the sample at 400 $^{o}$C in $N_2$ atmosphere, in order to promote the adhesion [17] of graphene to the nanoparticle film. We showed previously that the nanoscale rippling of graphene on NPs can be modified by annealing, and a small compressive strain can be induced [18]. Indeed, after annealing graphene follows more closely the morphology of the NPs, as shown by the AFM image in Figure 3a, where graphene covers the entire area.

Individual NPs can be very well distinguished. The corresponding adhesion map (Figure 3b) displays a variation of the adhesive force measured between the AFM tip and graphene. Comparing the adhesion map with the topography, it turned out that this variation is due to the condition in which graphene is present. Nanoparticle-supported graphene parts (e.g. blue circles in Figure 3a) display lower adhesion to the tip (blue circles in Figure 3b).

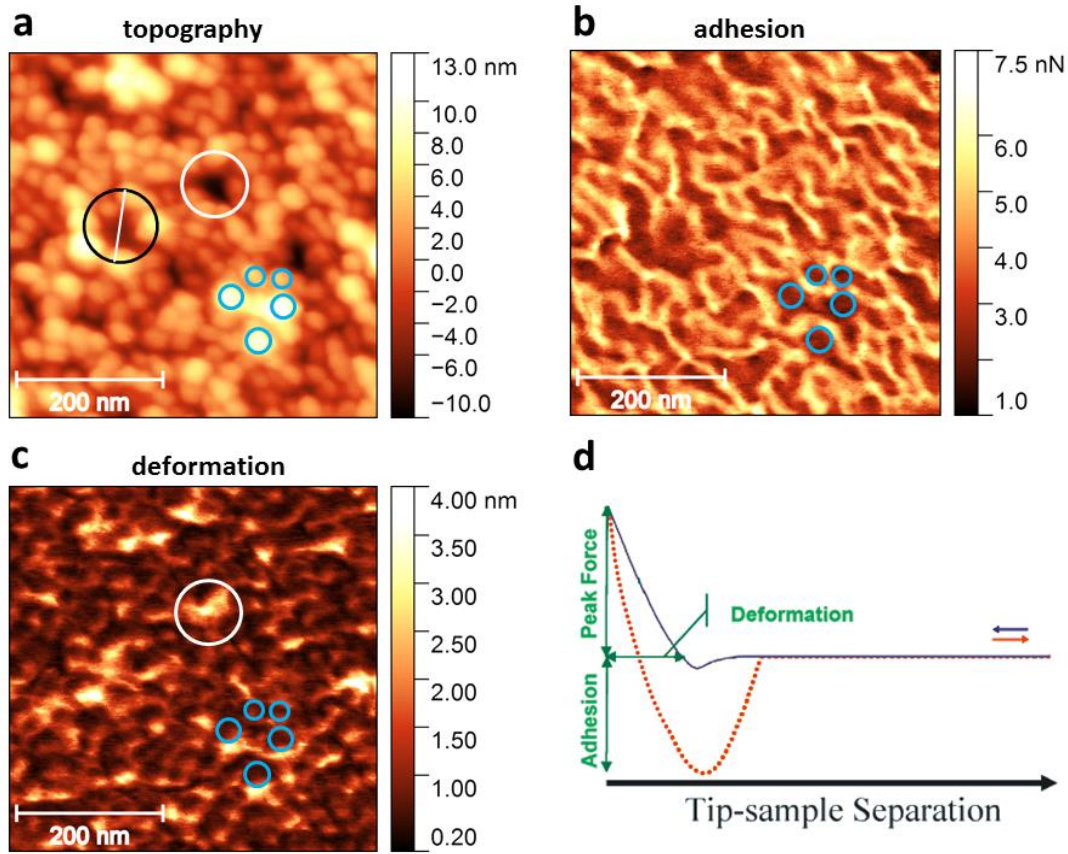

**Figure 3:** Simultaneously acquired Peak Force AFM images of graphene-covered nanoparticles after annealing. (a) Topography measured at a peak force of 8 nN. (b) Adhesion. (c) Deformation. (d) Typical force-distance curve (after Ref. [15]) indicating how adhesion and deformation data are extracted. Both the "approach" (blue line) and the "withdraw" (red dotted line) curves are displayed. Graphene is either directly supported by NPs (e.g. the areas denoted with blue circles in (a)-(c)) or it is suspended between them (e.g. the areas denoted with black (a) and white circles in (a) and (c)).

These graphene parts are not deformed significantly during the AFM measurements (blue circles in Figure 3c). In contrast, the graphene parts suspended between NPs display 2-3 times larger adhesion to the tip (light-coloured regions in Figure 3b). In some cases we observe suspended graphene areas which can be considered as circular membranes with diameters around 100 nm. Such suspended graphene membranes can be deformed by the AFM tip, as it can be observed in Figure 3c. Here, the larger deformation data inside the white circle corresponds to the suspended membrane marked also by white circle in Figure 3a. The deformation data in Figure 3c is calculated using the approach force-distance curve, and it is the difference in tip-sample separation from the point where the force is zero to the peak force point (Figure 3d). We performed nanoindentation experiments on the suspended graphene membrane in Figure 3a (black circle). The same area was scanned repeatedly by increasing gradually the peak force from 2 nN to 128 nN. A complete image was recorded for every force setpoint ($F$). Selected height profiles are shown in Figure 4a, which were taken along the same line section shown in Figure 3a (white line), extracted from the AFM images measured with the corresponding tip-sample force values. The force-induced deflection ($δ$) of the suspended graphene was measured as the difference between crests and troughs of the height profiles. Force-deflection data were obtained, as shown in Figure 4b.

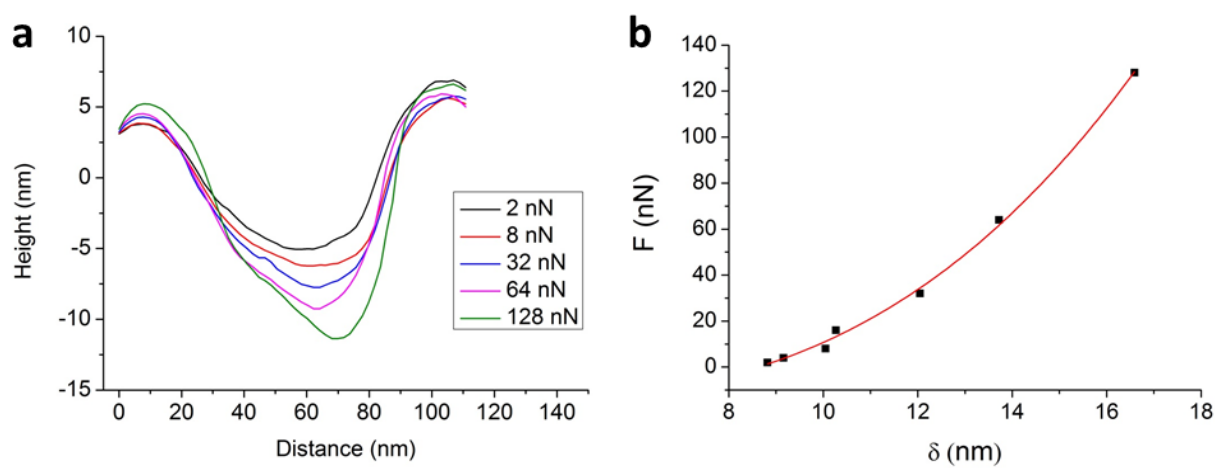

**Figure 4:** Nanoindentation performed on the suspended graphene nanomembrane marked with black circle in Figure 3a. (a) Height profiles taken along the same line section (white line in Fig. 3a), measured at different load forces ($F$). (b) Force-deflection data.

We note that the deflection induced by larger forces is reversible, and the indentation does not lead to plastic deformation of the suspended membrane. We interpret the experiments in the framework of the indentation model of a circular monolayer graphene using a spherical indenter [19]. The graphene area considered has a radius of approximately $a$ = 45 nm (Figure 3a, black circle), while the nominal radius of the AFM tip is $R \simeq 8$ nm. We fitted the force-deflection data with $F = c\delta + d\delta^3$, where the coefficients $c$ and $d$ are related to the Young's modulus $E$ and pre-tension $\sigma_0$ of a membrane of thickness $h$ (0.34 nm for graphene) [4, 19, 20, 21]. Since $R/a = 0.177$, we use the sphere load model ($R/a > 0.14$), with $d = Eq^3 a^{-2} h(R/a)^{1/4}$ [22, 23]. Here $q = 1/(1.05 - 0.15\nu - 0.16\nu^2) = 0.98$, with $\nu = 0.165$ the Poisson's ratio for monolayer graphene [4, 24]. From the fit we obtain the tensile modulus $E = 0.53$ TPa. We carried out similar nanoindentation experiments on other suspended graphene nanomembranes as well, using stiffer AFM cantilevers [18], and we calculated the tensile moduli as above. We obtained an average Young's modulus of $E_{\text{avg}} = 0.88$ TPa, which is close to the value of 1 TPa, found in previous experiments on both CVD-grown [6] and exfoliated samples [4, 25, 26]. The lower Young's modulus obtained in this work is probably related to the deviation of the actual geometry of the suspended graphene and the indenter from the circular and spherical shapes, respectively, used to fit the results. Another difference compared to previous works is that we used the sphere load model due to the higher $R/a$ ratio [19]. In this case the deviations from the ideal geometries can affect the result more significantly than in the case of point load model, where the radius of the graphene membrane is much larger than the radius of the indenter ($R/a < 0.03$) [4, 19]. Based on the AFM images performed at different forces, we do not observe any movement of the silica NPs during the indentation. However, we cannot exclude that graphene can slightly slide on some NPs when measured at larger forces, which can increase the measured deflection data and thus lower the derived Young's modulus.

**Conclusions**

In summary, we have investigated by Peak Force AFM the properties of CVD-grown graphene transferred onto a Langmuir-Blodgett film of $SiO_2$ nanoparticles. We showed by imaging at different interaction forces that the as-transferred graphene is only loosely bound to the nanoparticles, not following exactly their morphology. The binding of graphene to nanoparticles was improved by annealing.

We revealed extended graphene regions suspended between silica nanoparticles by comparing the simultaneously measured topography, adhesion, and deformation maps. The observed spatial variation in the adhesion is due to the fact that suspended graphene displays up to three times larger adhesion to the AFM tip. These graphene parts are also deformed more significantly during the AFM measurements. We investigated the elastic properties of suspended graphene by performing local indentation experiments on nanomembranes of 100 nm in diameter. We fitted the experimental force-deflection data using the sphere load model and obtained an average Young's modulus of $E_{avg} = 0.88$ TPa. Finally, we showed that the extrinsic morphology of transferred graphene is regulated by the underlying nanoparticles, which can open new pathways to fine tune the properties of graphene.

**Acknowledgments**

The research leading to these results has received funding from the People Programme (Marie Curie Actions) of the European Union's Seventh Framework Programme under REA grant agreement n° 334377, and from the Korea-Hungary Joint Laboratory for Nanosciences. The OTKA grants K-101599, PD-105173 in Hungary, as well as the János Bolyai Research Fellowships from the Hungarian Academy of Sciences are acknowledged. C. H.

acknowledges funding from the Nano-Material Technology Development Program (2012M3A7B4049888) through the National Research Foundation of Korea (NRF) funded by the Ministry of Science, ICT and Future Planning.**References**

1. A. K. Geim, "Graphene: status and prospects", Science, vol. 324, pp. 1530–1534, 2009.

2. I. A. Ovid'ko, "Mechanical properties of graphene", Reviews on Advanced Materials Science, vol. 34, pp. 1–11, 2013.

3. Ph. Lambin, "Elastic properties and stability of physisorbed graphene", Applied Sciences, vol. 4, pp. 282–304, 2014.

4. C. Lee, X. Wei, J.W. Kysar and J. Hone, "Measurement of the elastic properties and intrinsic strength of monolayer graphene", Science, vol. 321, pp. 385–388, 2008.

5. I. W. Frank, D. M. Tanenbaum, A. M. van der Zande and P. L. McEuen, Journal of Vacuum Science & Technology B, vol. 25, pp. 2558–2561, 2007.

6. G.-H. Lee, R. C. Cooper, S. J. An, S. Lee, A. van der Zande, N. Petrone, A. G. Hammerberg, C. Lee, B. Crawford, W. Oliver, J. W. Kysar and J. Hone, "High-strength chemical-vapor–deposited graphene and grain boundaries", Science, vol. 340, pp. 1073–1076, 2013.

7. M. Yamamoto, O. Pierre-Louis, J. Huang, M. S. Fuhrer, T. L. Einstein and W. G. Cullen, "«The princess and the pea» at the nanoscale: wrinkling and delamination of graphene on nanoparticles", Physical Review X, vol. 2, pp. 041018-1–041018-11, 2012.

8. T. Li, "Extrinsic morphology of graphene", Modelling and Simulation in Materials Science and Engineering, vol. 19, pp. 054005-1–054005-15, 2011.

9. M. I. Katsnelson and A. K. Geim, "Electron scattering on microscopic corrugations in graphene", Philosophical Transactions of the Royal Society A, vol. 366, pp. 195–204, 2008.
12